\pdfoutput=1

\documentstyle[pramana,floats,graphicx,amsmath]{ias}

\newcommand{\cP}{\mathcal{P}}
\newcommand{\cT}{\mathcal{T}}

\begin{document}

\title{Conduction bands in classical periodic potentials}

\author{Tanwa~Arpornthip${}^1$ and Carl~M~Bender${}^2$}

\address{Department of Physics, Washington University, St. Louis, MO 63130, USA
\\ ${}^1${\footnotesize{\tt email: tarpornthip@wustl.edu}}
\\ ${}^2${\footnotesize{\tt email: cmb@wustl.edu}}}

\abstract{The energy of a quantum particle cannot be determined exactly unless
there is an infinite amount of time in which to perform the measurement. This
paper considers the possibility that $\Delta E$, the uncertainty in the energy,
may be complex. To understand the effect of a particle having a complex energy,
the behavior of a classical particle in a one-dimensional periodic potential $V(
x)=-\cos(x)$ is studied. On the basis of detailed numerical simulations it is
shown that if the energy of such a particle is allowed to be complex, the
classical motion of the particle can exhibit two qualitatively different
behaviors: (i) The particle may hop from classically-allowed site to
nearest-neighbor classically-allowed site in the potential, behaving as if it
were a quantum particle in an energy gap and undergoing repeated tunneling
processes, or (ii) the particle may behave as a quantum particle in a conduction
band and drift at a constant average velocity through the potential as if it
were undergoing resonant tunneling. The classical conduction bands for this
potential are determined numerically with high precision.}

\maketitle

\section{Introduction}
\label{s1}

The theme of $\cP\cT$ quantum mechanics is that it is possible to extend the
Hamiltonian for a quantum system into the complex domain while retaining the
fundamental properties of a quantum theory \cite{R1,R2}. Complex quantum
mechanics has proved to be so interesting that the research activity on $\cP\cT$
quantum mechanics has motivated studies of complex classical mechanics. Early
work on the particle trajectories in complex classical mechanics is reported in
Refs.~\cite{R3,R4}. Subsequently, detailed studies of the complex extensions of
conventional classical-mechanical systems were undertaken: The remarkable
properties of complex classical trajectories were examined in
Refs.~\cite{R5,R6,R6a,R6b,R6c}; the complex behavior of the pendulum, the
Lotka-Volterra equations for population dynamics and the Euler equations for
rigid body rotation were studied in Refs.~\cite{R7,R8}; the complex
Korteweg-de Vries equation was examined in Refs.~\cite{R9,R10,R11,R12}; the
complex Riemann equation was examined in Ref.~\cite{R13}; complex Calogero
models were examined in Ref.~\cite{R14}; the complex extension of chaotic
behavior was recently discussed in Ref.~\cite{R15}.

This paper explores more deeply a newly discovered aspect of complex classical
mechanics; namely, that there are remarkable similarities between complex
classical mechanics and quantum mechanics: Bender, Brody, and Hook performed
numerical studies to show that standard quantum effects such as tunneling could
be displayed by classical systems having complex energy \cite{R16}. The work
in Ref.~\cite{R16} relies on the observation that when the energy is real, the
classical trajectories in the complex plane are closed and periodic but when the
energy is complex, the classical trajectories are open and nonperiodic
\cite{R8}.

\begin{figure*}[t!]
\begin{center}
\includegraphics[scale=0.95, viewport=0 0 200 170]{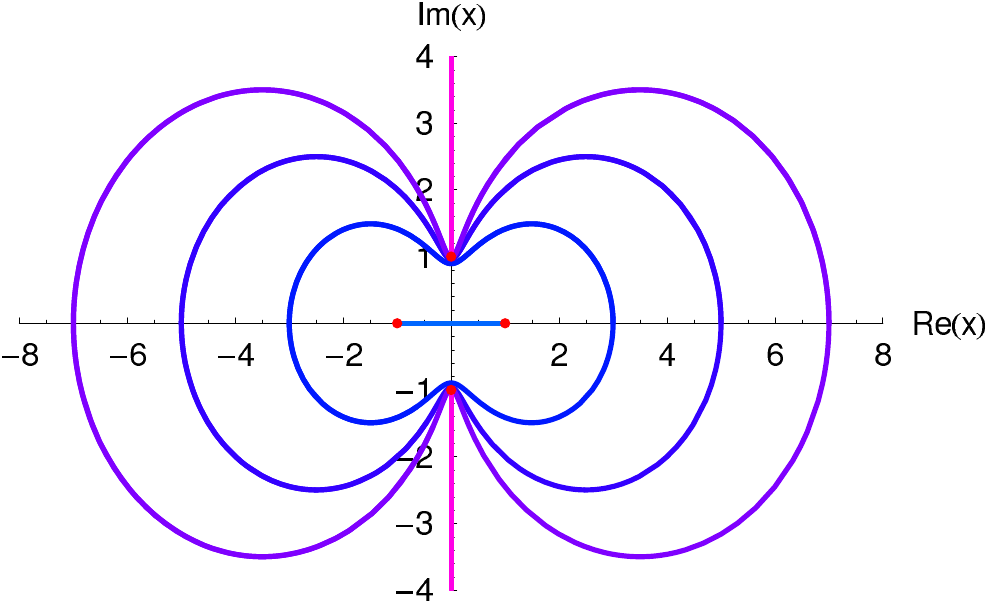}
\end{center}
\caption{Classical trajectories $x(t)$ in the complex-$x$ plane for the
anharmonic-oscillator Hamiltonian $H=\frac{1}{2}p^2+x^4$. All trajectories
represent a particle of energy $E=1$. There is one real trajectory that
oscillates between the turning points at $x=\pm1$ and an infinite family of
nested complex trajectories that enclose the real turning points but lie inside
the imaginary turning points at $\pm i$. (The turning points are indicated by
dots.) Two other trajectories begin at the imaginary turning points and drift
off to infinity along the imaginary-$x$ axis. Apart from the trajectories
beginning at $\pm i$, all trajectories are closed and periodic. All orbits in
this figure have the same period $\sqrt{\pi/2}\,\Gamma\left(\frac{1}{4}\right)/
\Gamma\left(\frac{3}{4}\right)=3.70815\ldots$.}
\label{f1}
\end{figure*}

In Ref.~\cite{R16} these features are illustrated by using the simple example of
the classical anharmonic oscillator, whose Hamiltonian is $H=\frac{1}{2}p^2+
x^4$. When the classical energy is real, the classical orbits in the complex
plane are all closed and periodic (see Fig.~\ref{f1}), but if the energy is
complex, the classical orbits are no longer closed and they spiral outward to
infinity (see Fig.~2).

\begin{figure*}[t!]
\begin{center}
\includegraphics[scale=0.95, viewport=0 0 200 145]{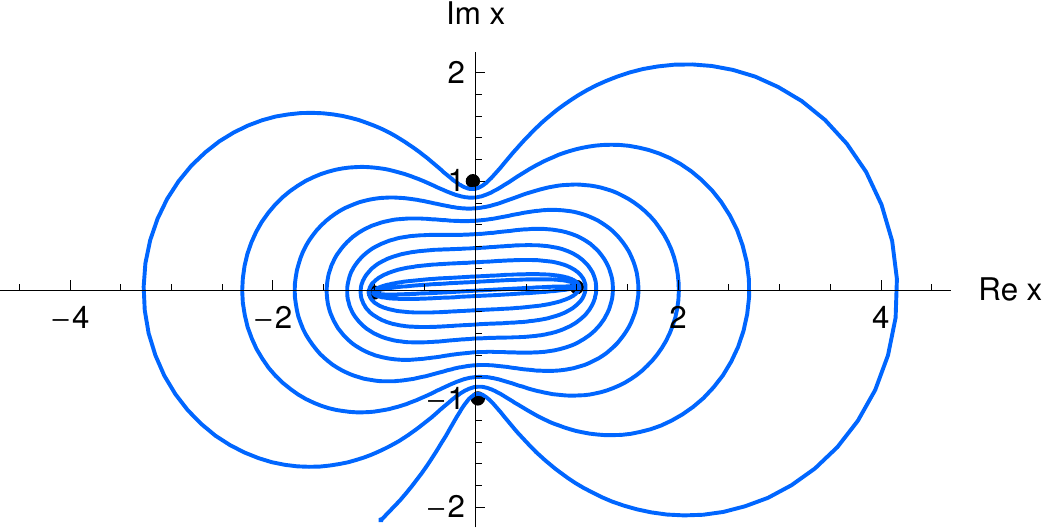}
\caption{A single classical trajectory in the complex-$x$ plane for a particle
governed by the anharmonic-oscillator Hamiltonian $H=\frac{1}{2}p^2+x^4$. This
trajectory begins at $x=1$ and represents the complex path of a particle whose
energy $E=1+0.1i$ is complex. The trajectory is not closed or periodic. The four
turning points are indicated by dots. The trajectory does not cross itself.}
\end{center}
\label{f2}
\end{figure*}

The Bohr-Sommerfeld quantization formula provides a heuristic way to understand
the association between real energy and closed orbits versus complex energy and
open orbits. The quantization formula
\begin{equation}
\oint p dx\sim\left(n+\textstyle{\frac{1}{2}}\right)\pi\quad(n>>1),
\label{e1}
\end{equation}
where the integral is performed along a closed classical orbit, gives a
semiclassical approximation to the energies of a Hamiltonian for large quantum
number; that is, it takes advantage of particle-wave duality and assumes that
there are an integral number of wavelengths along the classical orbit. For the
anharmonic-oscillator Hamiltonian considered in Figs.~\ref{f1} and 2 this
formula gives the (real) spectrum of this Hamiltonian, whether or not the
classical orbit lies on the real axis in the complex-$x$ plane. However, the
formula cannot be applied if the orbit is open and the energy is complex.

\begin{figure*}[t!]
\begin{center}
\includegraphics[scale=0.95, viewport=0 0 200 175]{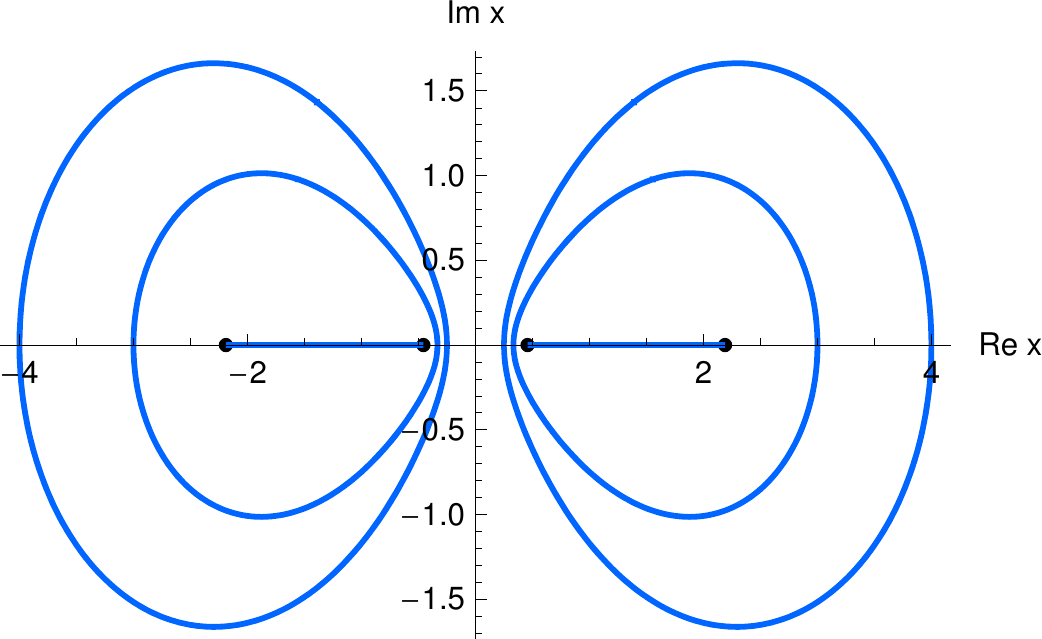}
\end{center}
\caption{Six classical trajectories in the complex-$x$ plane representing a
particle of energy $E=-1$ in the potential $x^4-5x^2$. The turning points are
located at $x=\pm2.19$ and $x=\pm0.46$ and are indicated by dots. Because the
energy is real, the trajectories are all closed. The classical particle stays in
either the right-half or the left-half plane and cannot cross the imaginary
axis. Thus, when the energy is real, there is no effect analogous to tunneling.}
\label{f3}
\end{figure*}

In Ref.~\cite{R16} it was seen that it is possible to observe tunneling-like
behavior in classical mechanics when the classical energy of a particle is taken
to be complex. The double-well anharmonic oscillator $H=p^2+x^4-5x^2$ was used
to illustrate this observation. If the energy of a classical particle is taken
to be real, the classical motion is periodic and the orbits lie either on
the left side or on the right side of the imaginary axis, which separates
the two wells (see Fig.~\ref{f3}).

\begin{figure*}[t!]
\begin{center}
\includegraphics[scale=0.95, viewport=0 0 200 280]{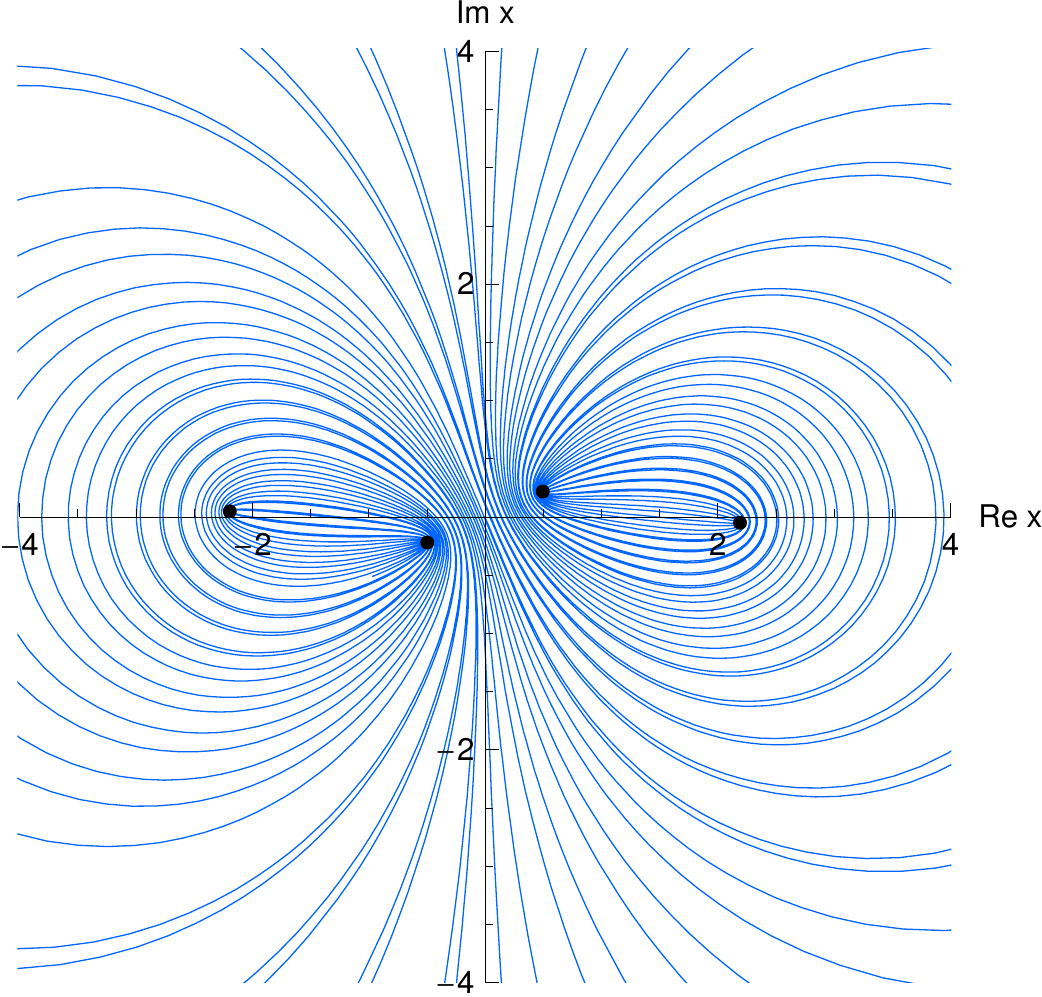}
\end{center}
\caption{Classical trajectory of a particle moving in the complex-$x$ plane
under the influence of a double-well $x^4-5x^2$ potential. The particle has
complex energy $E=-1-i$ and thus its trajectory does not close. The trajectory
spirals outward around one pair of turning points, crosses the imaginary axis,
and then spirals inward around the other pair of turning points. It then spirals
outward again, crosses the imaginary axis, and goes back to the original pair of
turning points. The particle repeats this behavior endlessly but at no point
does the trajectory cross itself. This classical-particle motion is analogous to
the behavior of a quantum particle that repeatedly tunnels between two
classically allowed regions. Here, the particle does not disappear into the
classically forbidden region during the tunneling process; rather, it moves
along a well-defined path in the complex-$x$ plane from one well to the other.}
\label{f4}
\end{figure*}

However, if the energy of a classical particle whose motion is determined by the
Hamiltonian $H=p^2+x^4-5x^2$ is taken to be complex, it was found that the
classical trajectory is no longer closed. Rather, the trajectory spirals
outward, crosses the imaginary axis, and then spirals {\it inward} into the
companion well. It then repeats this process, alternately visiting one well and
then the other. This motion is shown in Fig.~\ref{f4}. Note that the trajectory
never crosses itself. This strange-attractor-like behavior resembles
quantum-mechanical tunneling. Moreover, it provides a heuristic answer to the
question, How did the particle get from one well to the other? In a tunneling
experiment the particle disappears from one well and then appears in the other
well; what path did the particle follow to get from one well to the other?
Figure \ref{f4} is that the particle can travel smoothly from one well to the
other if the trajectory lies in the complex plane.

The heuristic picture proposed in Ref.~\cite{R16} is that there is some quantum
uncertainty in the measurement of the energy of a particle, $(\Delta E)(\Delta
t)\sim\hbar$, because an energy measurement is performed in a finite, rather
than an infinite, amount of time. If this uncertainty in the energy has an
imaginary component, then the classical picture of particle motion and the
quantum picture of particle motion come closer to one another.

The question that is investigated in this paper is how a classical particle
whose energy is allowed to be complex behaves in a {\it periodic} potential. In
Ref.~\cite{R16} some numerical experiments were performed for the potential $V(x
)=-\cos(x)$ which suggested that depending on the real and the imaginary parts
of the energy of a particle one can observe two kinds of classical motion:
\begin{itemize}
\item{} Localized tunneling behavior -- The particle tunnels from well to
adjacent well, either to the right or to the left. This hopping behavior
resembles a deterministic random walk, and is what is usually observed for an
arbitrary choice of complex energy. This motion is like that of a quantum
particle in a crystal, where the energy of the particle is not in a conduction
band.
\item{} Delocalized conduction -- The particle drifts through the potential in
one direction. This classical motion resembles the behavior of a quantum
particle in a crystal, where the energy of the particle lies in a conduction
band.
\end{itemize}

These two kinds of behaviors were conjectured in Ref.~\cite{R16} on the basis of
limited numerical experiments. We have now performed an extremely detailed
numerical study of the behavior of a classical particle in a periodic potential
well. Our study shows that the types of behavior that were conjectured in
Ref.~\cite{R16} do indeed occur. More importantly, there are well-defined and
narrow energy bands with precise edges in which ``conduction" takes place.

\section{Complex motion in a periodic potential}
\label{s2}

\begin{figure*}[bth]
\begin{center}
\includegraphics[scale=0.40, viewport=0 0 600 450]{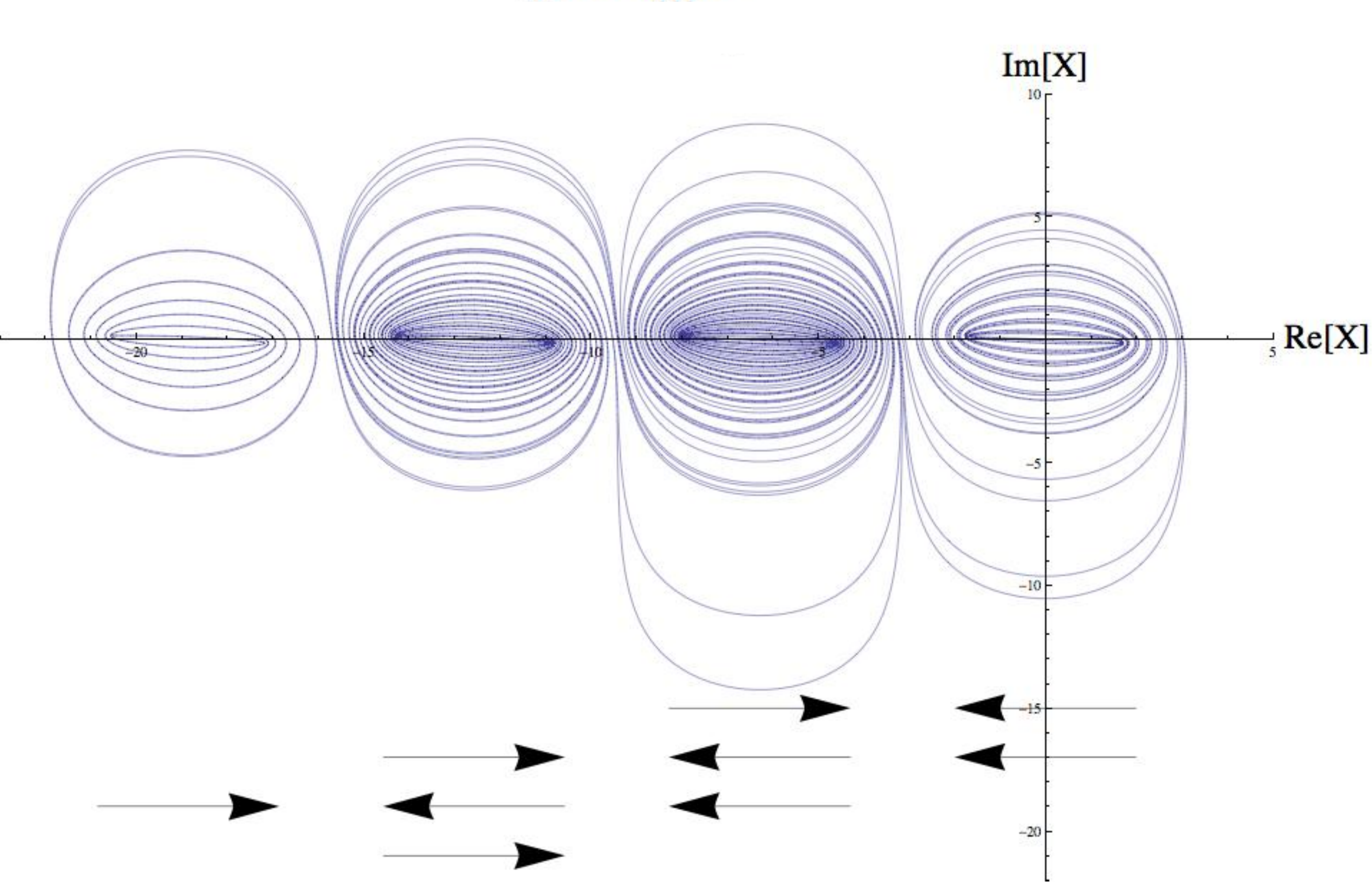}
\end{center}
\caption{A tunneling trajectory for the Hamiltonian (\ref{e2}) with $E=0.1-0.15
i$. The classical particle hops at random from well to well in a random-walk
fashion. The particle starts at the origin and then hops left, right, left,
left, right, left, left, right, right. This is the sort of behavior normally
associated with a particle in a crystal at an energy that is not in a
conduction band. At the end of this simulation the particle is situated to the
left of its initial position. The trajectory never crosses itself.}
\label{f5}
\end{figure*}

\begin{figure*}[t!]
\begin{center}
\includegraphics[scale=0.40, viewport=0 0 800 585]{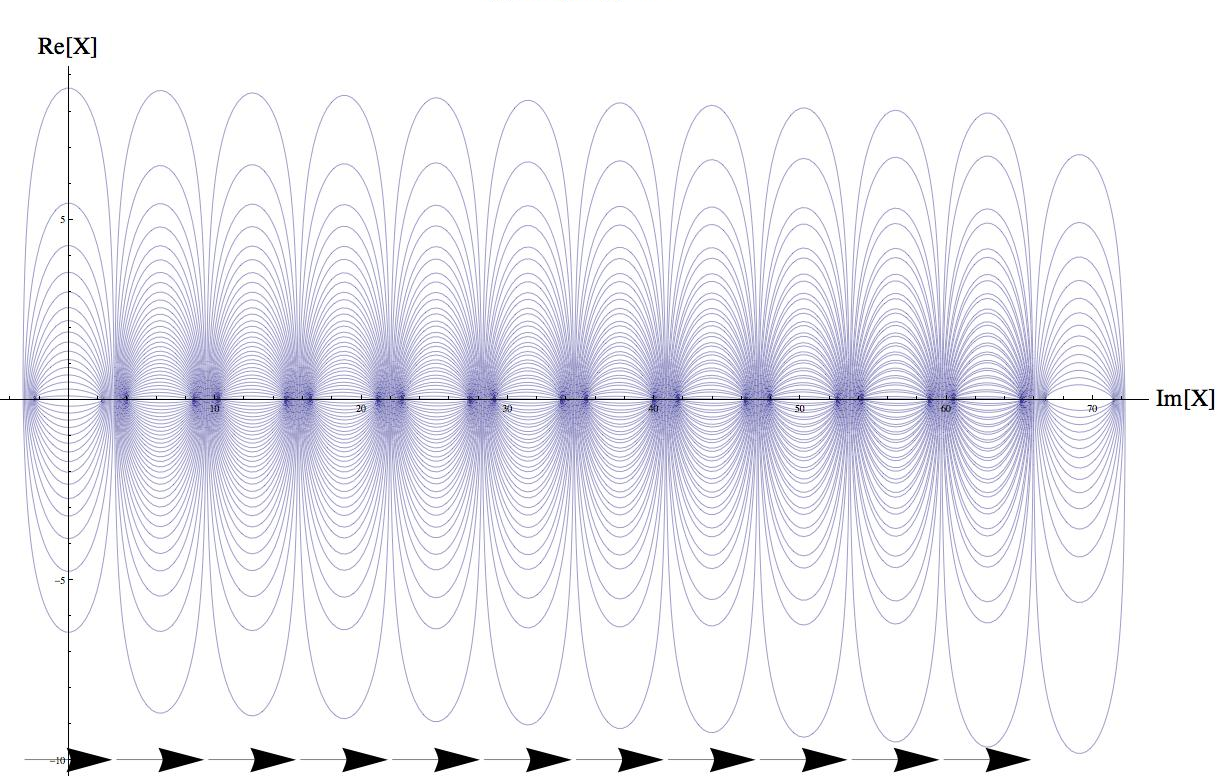}
\end{center}
\caption{A classical particle exhibiting a behavior analogous to that of a
quantum particle in a conduction band that is undergoing resonant tunneling.
Unlike the particle in Fig.~\ref{f5}, this classical classical particle tunnels
in one direction only and drifts at a constant average velocity through the
potential.}
\label{f6}
\end{figure*}

We consider here the Hamiltonian
\begin{equation}
H(x,p)=\textstyle{\frac{1}{2}}p^2-\cos(x)
\label{e2}
\end{equation}
and solve the associated Hamilton's equations of motion,
\begin{eqnarray}
\frac{dx}{dt}=p\quad{\rm and}\quad
\frac{dp}{dt}=-\sin(x),
\label{e3}
\end{eqnarray}
to determine the kinds of classical behaviors that arise. If we choose a value
of the classical energy at random, say $E=0.1-0.15i$, we find that the classical
particle usually tunnels from well to well following a deterministic random walk
(see Fig.~\ref{f5}). On the other hand, if we search carefully for a conduction
band, we see the classical particle drift through the lattice in one direction
only (see Fig.~\ref{f6}).  

\begin{figure*}[bth]
\includegraphics[scale=0.50, viewport=0 0 600 540]{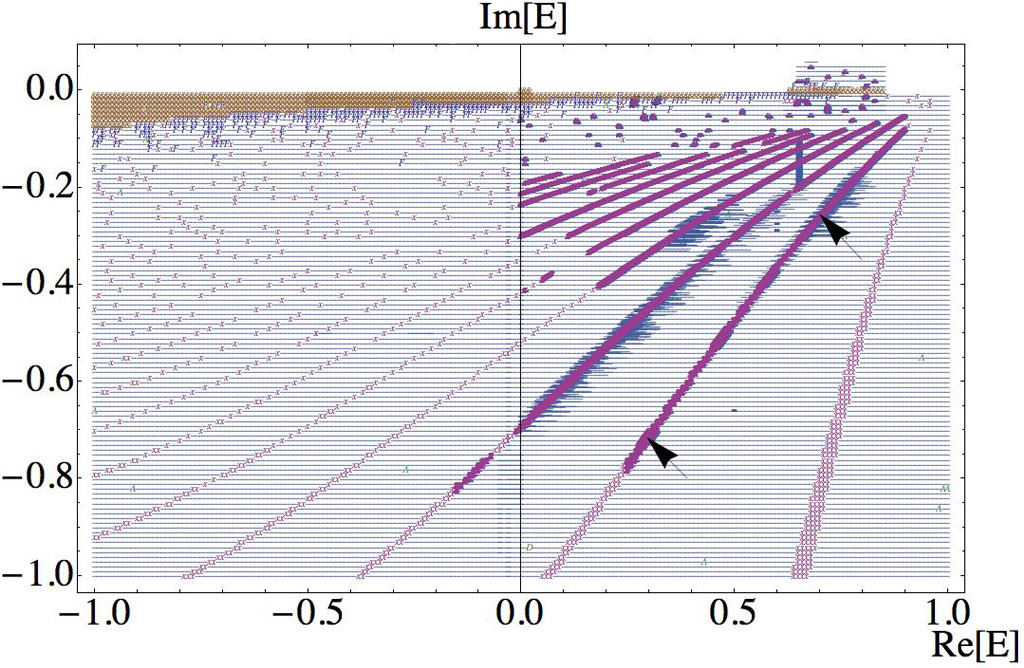}
\caption{Complex-energy plane showing those energies that lead to tunneling
(hopping) behavior and those energies that give rise to conduction. Hopping
behavior is indicated by a hyphen - and conduction is indicated by an X. The
symbol \& indicates that no tunneling takes place; tunneling does not occur for
energies whose imaginary part is close to 0. In some regions of the energy plane
we have done very intensive studies and the X's and -'s are densely packed. This
picture suggests the features of band theory: If the imaginary part of the
energy is taken to be $-0.9$, then as the real part of the energy increases from
$-1$ to $+1$, five narrow conduction bands are encountered. These bands are
located near ${\rm Re}\,E=-0.95,\,-0.7,\,-0.25,\,0.15,\,0.7$. This picture is
symmetric about ${\rm Im}\,E=0$ and the bands get thicker as $|{\rm Im}\,E|$
increases. A total of 68689 points were classified to make this plot. In most
places the resolution (distance between points) is $0.01$, but in several
regions the distance between points is shortened to $0.001$. The regions
indicated by arrows are blown up in Figs.~\ref{f8} and \ref{f9}.}
\label{f7}
\end{figure*}

We have run continuously on a computer for several months to determine which
complex energies give rise to tunneling (hopping) behavior and which complex
energies produce conduction-like behavior. We have been able to determine that
conduction behavior is found when the energy lies in narrow bands with crisp
well-defined edges (see Fig.~\ref{f7}). To distinguish between hopping behavior
and conduction behavior we allowed the particle to undergo 10 tunneling
incidents, and if the particle always tunneled in the same direction we
classified the energy as lying in a conduction band. (In Ref.~\cite{R16}, seven
tunneling incidents were used.) Figure \ref{f7} reports the major result in this
paper.

\begin{figure*}[bth]
\begin{center}
\includegraphics[scale=0.30, viewport=0 0 850 450]{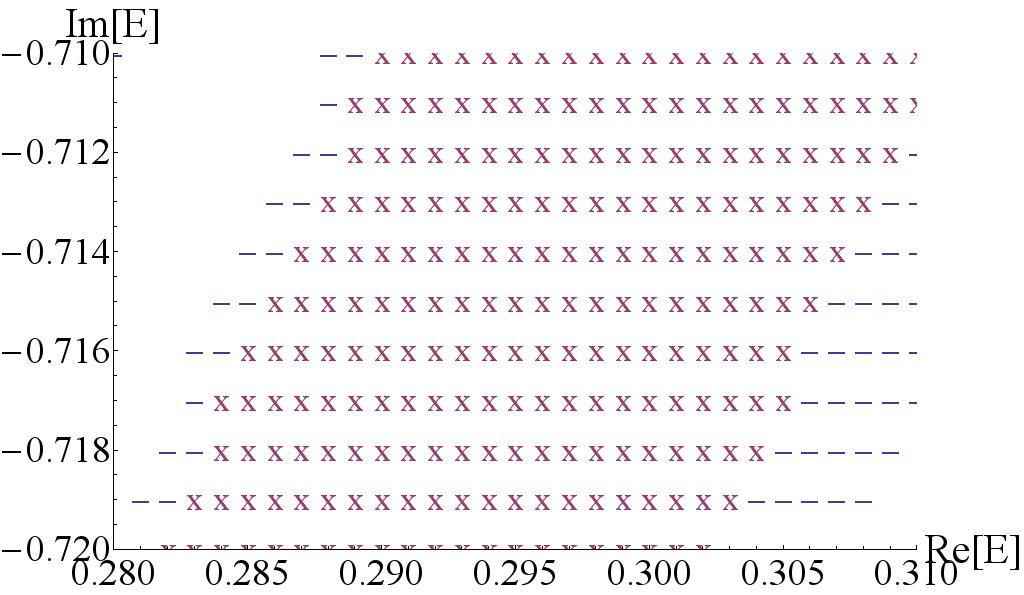}
\end{center}
\caption{Detailed portion of the complex-energy plane shown in Fig.~\ref{f8}
containing a conduction band. Note that the edge of the conduction band, where
tunneling (hopping) behavior changes over to conducting behavior, is very
sharp.}
\label{f8}
\end{figure*}

\begin{figure*}[bth]
\begin{center}
\includegraphics[scale=0.30, viewport=0 0 850 450]{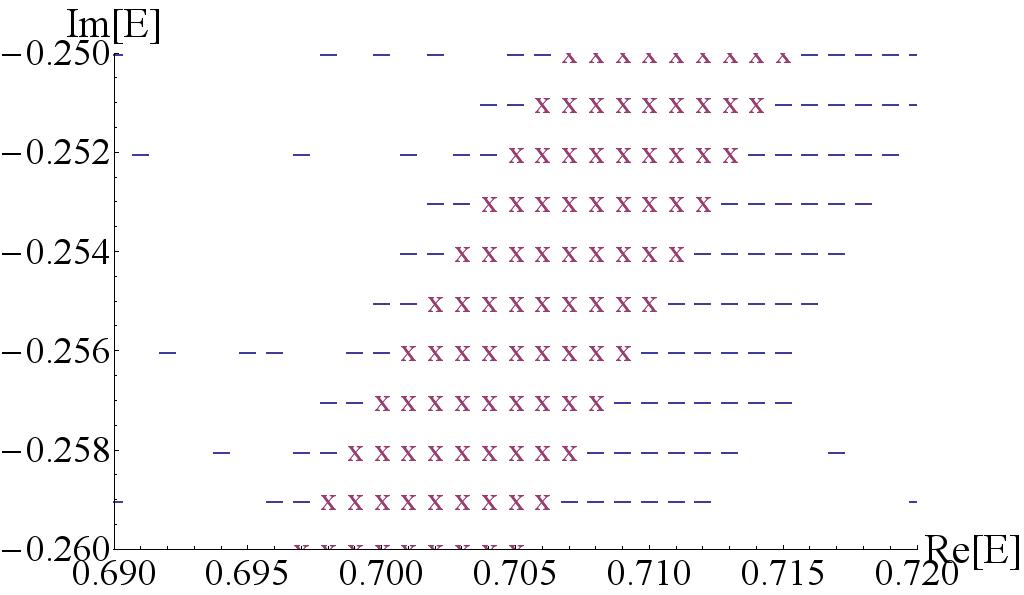}
\end{center}
\caption{Like Fig.~\ref{f8}, a detailed portion of the complex-energy plane in
Fig.~\ref{f8} containing a conduction band. Note that the edges of the
conduction band are sharp.}
\label{f9}
\end{figure*}

Two detailed blow-up views of the complex-energy plane shown in Fig.~\ref{f7}
are given in Figs.~\ref{f8} and \ref{f9}. In Fig.~\ref{f8} a region of the
complex-$E$ plane is shown for which $0.28<{\rm Re}\,E<0.31$ and $-0.72<{\rm
Im}\,E<-0.71$, and in Fig.~\ref{f9} a portion of the complex-$E$ plane is shown
for which $.69<{\rm Re}\,E<0.72$ and $-0.26<{\rm Im}\,E<-0.25$. These
figures indicate that the border between hopping behavior and conduction
behavior is clean and crisp.

\section{Concluding remarks}
\label{s3}
\begin{figure*}[bth]
\begin{center}
\includegraphics[scale=0.70, viewport=0 0 350 200]{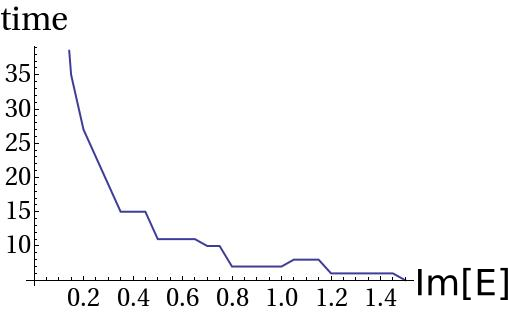}
\end{center}
\caption{Tunneling time as a function of the imaginary part of the energy. The
curve is a hyperbola; that is, the product of the imaginary part of the energy
and the tunneling time is a constant. This result is strongly reminiscent of the
time-energy uncertainty principle.}
\label{f10}
\end{figure*}

We know that quantum-mechanical amplitudes are determined by performing infinite
sums over classical configurations. However, tunneling is not thought of as a
possible classical behavior. Thus, the usual view is that quantum phenomena 
can appear only when an infinite sum over classical behaviors is performed.
However, in this paper we have argued that some aspects of quantum behavior
(such as bands and gaps in periodic potentials) can already be seen in the
context of classical mechanics. As an example, we can already see in classical
mechanics a version of the time-energy uncertainty principle. In Fig.~\ref{f10}
we have plotted the tunneling time versus the size of the imaginary part of
the energy. The graph is a hyperbola, and thus the product of these two
quantities is a constant.

\vspace{0.5cm}
\footnotesize
\noindent
We thank D.~Hook for many interesting discussions. CMB is supported by a grant
from the U.S.~Department of Energy.
 
\end{document}